\documentstyle[emulateapj,onecolfloat,psfig]{article}

\newcommand{\da}{d_A}

\newcommand{\veck}{{\bf k}}
\newcommand{\vecl}{{\bf l}}

\newlength{\tskip}
\newlength{\colwidth}

\newcommand{\beq}{\begin{equation}}
\newcommand{\eeq}{\end{equation}}
\newcommand{\beqa}{\begin{eqnarray}}
\newcommand{\eeqa}{\end{eqnarray}}
\begin{document}
\twocolumn[
\title{Free-Free Emission at Low Radio Frequencies}
\author{Asantha Cooray and Steven R. Furlanetto}
\affil{Theoretical Astrophysics, Division of Physics, Mathematics and
  Astronomy,  
California Institute of Technology, MS 130-33, Pasadena, CA 91125\\}

\begin{abstract}
We discuss free-free radio emission from ionized gas in the
intergalactic medium.  Because the emissivity is proportional to the
square of the electron density, the mean background is strongly
sensitive to the spatial clumping of free electrons.  Using several
existing models for the clumping of ionized gas, we find that the
expected free-free distortion to the cosmic microwave background (CMB)
blackbody spectrum is at a level detectable with upcoming experiments
such as the Absolute Radiometer for Cosmology, Astrophysics, and
Diffuse Emission (ARCADE).  However, the dominant contribution to the
distortion comes from clumpy gas at $z \la 3$, and the integrated
signal does not strongly constrain the epoch of reionization.  In
addition to the mean emission, we consider spatial fluctuations in the
free-free background and the extent to which these anisotropies
confuse the search for fluctuations in 21 cm line emission from
neutral hydrogen during and prior to reionization.  This background is
smooth in frequency space and hence can be removed through frequency
differencing, but only so long as the 21 cm signal and the free-free
emission are uncorrelated.  We show that, because the free-free
background is generated primarily at low redshifts, the
cross-correlation between the two fields is smaller than a few
percent.  Thus, multifrequency cleaning should be an effective way to
eliminate the free-free confusion.
\end{abstract}

]

\section{Introduction}

Plans for upcoming low-frequency radio experiments able to measure the
21 cm background associated with neutral hydrogen during, and prior
to, reionization era (e.g., \cite{Scott90} 1990; \cite{Madetal97}
1997; \cite{Zaletal03} 2003; \cite{MorHew03} 2003) have motivated
study of the foregrounds that may contaminate such measurements.
Synchrotron emission from the Milky Way, low frequency radio sources
(\cite{DiMetal02} 2002), and free-free emission from free electrons in
the intergalactic medium (\cite{OhMac03} 2003) are now thought to be
the chief sources of confusion.

While the free-free background contaminates 21 cm studies, the
emission itself captures important physics of the ionized component of
the intergalactic medium (IGM). In particular, because the emissivity
is proportional to the square of the electron number density,
free-free emission is strongly sensitive to whether electrons are
spatially clumped or distributed smoothly in the IGM. Initial
estimates of the free-free background suggest that gas clumping boosts
the specific intensity by a factor from order unity at $z\sim20$ to
over 100 at $z \la 3$ (\cite{Oh99} 1999).

Here, we consider the direct detection of free-free emission through
the distortion it creates in the cosmic microwave background (CMB)
blackbody spectrum at low radio frequencies (\cite{Loeb96} 1996;
\cite{Oh99} 1999).  We calculate the mean distortion temperature using
a variety of both analytic and numeric clumping models already
existing in the literature.  While the expected brightness varies by
factors of a few between the different models, we suggest that the
distortion to the CMB is at a level detectable with the planned
Absolute Radiometer for Cosmology, Astrophysics, and Diffuse Emission
(ARCADE; \cite{Kog03} 20003) experiment.  
%SF
A detection of the distortion could help to discriminate between
existing clumping models and to constrain the integrated ionization
history of the universe.
%SF, end
We also show that ionized halos at $z \la 3$ dominate the background,
with only a relatively insignificant contribution from the
reionization era.

In addition to the mean background, we estimate the magnitude of spatial
fluctuations in the free-free intensity that appear because the
ionized clumps are a biased tracer of the dark matter density field.
We find that the fluctuations are comparable to or greater than those
from the 21 cm signal on the relevant scales.  Fortunately, as shown
by \cite{Zaletal03} (2003), the smoothness of the free-free background
in frequency space should allow one to clean these sources in 21 cm maps
(see also \cite{MorHew03} 2003); the cleaning is quite efficient but
relies on the foreground and 21 cm signals being uncorrelated.  In
fact the two should be anticorrelated, because the free-free emission
comes from ionized halos while the 21 cm signal comes from neutral
regions.  Here we show that, because nearly all of the free-free
background arises at $z \la 3$, the anti-correlation is at the level
of few percent, suggesting that adequate cleaning is possible.
%As a result, free-free fluctuations
%can be removed from 21 cm fluctuations based on multifrequency
%information, to the extent that free-free fluctuations are correlated
%with 21 cm fluctuations.  Given the small overlap in redshift space,
%this cross- correlation is 
%Furthermore, arcminute scale anisotropy information of the free-free background, at
%frequencies of order few GHz, can be used to clean the 100 MHz level
%contamination. While cleaning the few GHz data requires  
%that dominant fluctuations related to intrinsic anisotropies in CMB
%be removed, this can easily be achieved with  
%either existing information from experiments such as WMAP or upcoming
%data such as Planck, which produce 
%CMB anisotropy maps limited by cosmic variance down to tens or
%arcminute scales and below. 

%Compared to previous calculations (e.g., \cite{OhMac03} 2003), our
%conclusions are some what different. While we suggest 
%that the free-free background is interesting in its own right for
%astrophysical purposes, mainly to understand the 
%clumping of free electrons in the IGM, we also suggest that its
%confusion can be significantly reduced for future 21 cm measurements.  
The discussion is organized as follows: in \S 2, we briefly discuss
free-free emission from ionized halos and its potential detectability
as a distortion to the CMB.  In \S 3, we discuss the detectability of
spatial fluctuations in the free-free background and the extent to
which free-free emission contaminates 21 cm measurements. Throughout
the paper, we make use of the WMAP-favored LCDM cosmological model
(\cite{Speetal03} 2003).

\begin{figure}[t]
\psfig{file=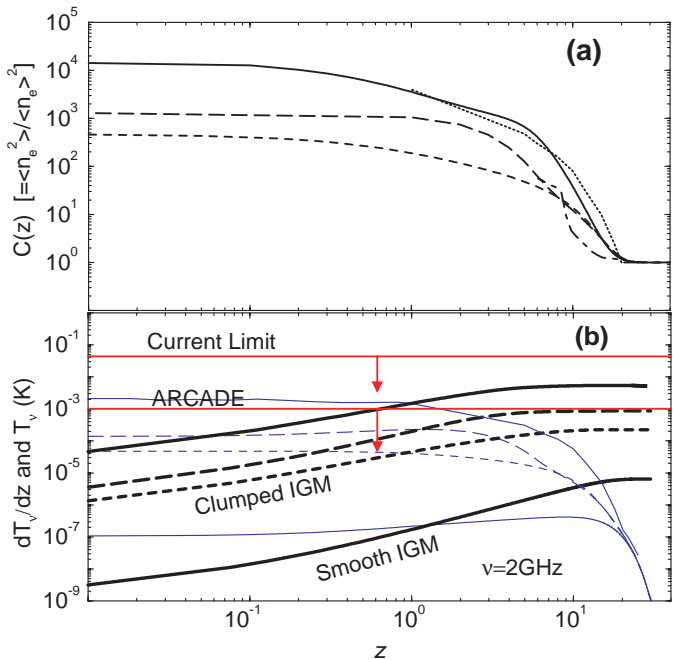,width=3.5in,angle=0}
\caption{(a) The clumping factor of electrons in the IGM as a function
of redshift.  These clumping factors come from \cite{Haietal01} (2001)
(long-dashed line: a reionization model that includes minihalos as
well as star-forming halos with $T_v > 10^4$~K), \cite{Benetal01}
(2001) (dashed line: semi-analytic model including cool neutral gas,
or galaxies, in ionized halos as well as the clumping of gas outside
halos), and \cite{GneOst97} (1997) (dot-dashed line: a direct
measurement in numerical simulations).  The dotted line is the
clumping factor presented in \cite{Oh99} (1999), while the solid line
is a similar calculation described in the text based on the
star formation history of the universe.  (b) The mean brightness
temperature of free-free emission at an observed frequency of 2 GHz.
Here, we show both the differential (thin lines) and cumulative
brightness temperatures (thick lines) as a function of the
redshift. The curves refer to the clumping models in the top panel.
For reference, we show the case where $C(z)=1$ and the ionized gas
distribution is taken to be spatially smooth.  We also show the
current limit on the free-free distortion of the CMB (top line with
$Y_{\rm ff} \equiv T_\nu/T_{\rm CMB} (h\nu/kT_{\rm CMB})^2 < 1.9
\times 10^{-5}$~K; \cite{Beretal94} 1996) and the expected constraint
from the ARCADE experiment (\cite{Kog03} 2003).}
\label{fig:clump}
\end{figure}

\section{The Mean Free-Free Signal}

First, we write the free-free emission coefficient as
\begin{eqnarray}
\epsilon_\nu &=& 5.4 \times 10^{-39} n_e^2 T_e^{-1/2} g_{ff}(\nu,T_e) e^{-h\nu/kT_e} \nonumber \\
&& \quad \quad \quad \, {\rm ergs \; cm^{-3}\; s^{-1}\; Hz^{-1}\; sr^{-1}} \, ,
\label{eqn:free}
\end{eqnarray}
where the Gaunt-factor can be approximated in the radio regime as
$g_{ff}(\nu,T_e) \approx 11.96 T_e^{0.15} \nu^{-0.1}$ (\cite{Lan99}
1999).  The cumulative specific intensity is simply $I_\nu = \int
d\chi \epsilon_\nu(z) x_e^2(z) \\ C(z)/(1+z)^4$, where $\chi$ is the
conformal time or lookback distance from the observer and $\nu(z) =
\nu_{\rm obs}(1+z)$ with $\nu_{\rm obs}$ the observed frequency.  The
electron fraction $x_e(z)$ captures the ionization history of the
IGM. In our calculations, unless specified as part of the clumping
description, we take a model for reionization based on the UV
radiation from star formation (e.g., \cite{HaiHol03} 2003;
\cite{Cheetal03} 2003), with a cumulative optical depth to electron
scattering of $\sim$ 0.14 (consistent with the recent measurements
from CMB polarization data; \cite{Kogetal03} 2003).  We will quote
results in terms of the brightness temperature $T_b = c^2
I_\nu/2\nu^2_{\rm obs}k_B$ (valid for the Rayleigh-Jeans part of the
spectrum).

In Eq.~(\ref{eqn:free}), we have introduced $C(z) = \langle n_e^2
\rangle/\langle n_e \rangle^2$, the clumping factor of the electron
density field.  As has been noted previously (\cite{Oh99} 1999), the
free-free intensity depends strongly on this factor. 
Instead of choosing a specific prescription, we make use of a variety
of published models for $C(z)$.  One approach is to estimate the
clumping from the large-scale matter distribution (by, for example,
assuming that ionized electrons reside in virialized dark matter
halos).  Several such models are shown in Fig.~1(a), including
estimates based on analytic (\cite{Haietal01} 2001), semi-analytic
(\cite{Benetal01} 2001), and numerical (\cite{GneOst97} 1997)
techniques.  These clumping models differ in detail, especially
around $z \sim 10$, because they include different aspects of the
reionization process, such as the presence of partly ionized minihalos
(\cite{Haietal01} 2001).  

The second approach, taken by \cite{Oh99} (1999), is to construct a
model for the production rate of ionizing photons $\dot{n}_{\rm ion}$
and assume that, throughout the universe, ionization equilibrium is a
good approximation.  The quantity $x_e^2 C(z)$ is then fixed through
the relation $\dot{n}_{\rm ion} = x_e^2 (z)n_e^2(z) C(z) \alpha_{\rm
rec}$, where $\alpha_{\rm rec}$ is the recombination coefficient.  We
show the clumping factor estimated by \cite{Oh99} (1999) in Fig. 1(a)
as a dotted line; at high redshifts it assumes that the ionizing
photon production rate is proportional to the mass in halos with
virial temperatures $T_v > 10^4$~K (with $\sim 400$ ionizing photons
per collapsed baryon, appropriate for a star formation efficiency of
$10\%$ and Population II stars with a standard IMF).  We also show a
similar estimate using the analytical model of \cite{HerSpr03} (2003)
for the star formation history of the universe, again assuming
Population II stars with a standard IMF.  While the star formation
history is model dependent at high redshifts, this model fits
observational measurements at $z \la 3$ (e.g., \cite{Som01} 2001)
reasonably well.  The \cite{Oh99} (1999) approach predicts stronger
clumping than models based on models of the IGM gas do, suggesting
that the latter approach may underestimate the clumpiness of galaxies.
However, all of the clumping models generally have similar
evolutionary histories, with $C(z=20) \sim 1$ increasing to $C(z=0)
\sim 10^3$.  We will find below that the detailed behavior at
$z\sim10$ has little effect on the observable results.

As shown in Fig.~1(b), $dT_\nu/dz$ closely traces the redshift
evolution of $C(z)$.  In a smooth IGM, with $C(z)=1$, a wide range of
redshifts contribute approximately equally to the free-free
background, decreasing only around reionization ($z \sim 15$ in our
model).  Clumping increases the contribution to the free-free
background by a factor $C(z)$ at each redshift: because of the sharp
increase in the clumping factor at $z \sim 3$, most of the free-free
background arises at these low redshifts and not during the
reionization epoch.  Unfortunately, since one measures only the
integrated background, the redshift evolution of the clumping factor
is not directly observable.  Note that the models considered give
predictions that vary by a factor of a few.  This suggests that these
measurements may provide a discriminant between some of these models,
especially between the two different approaches to clumping
calculations.  We also see that, independent of the detailed modeling,
the predicted distortion to the CMB spectrum is at a level that can be
detected with ongoing experiments, such as ARCADE (\cite{Kog03} 2003).
While such a detection would not allow us to extract with confidence a
specific model for $C(z)$ or to constrain the reionization history
independently, a detection or upper limit of $\sim 10^{-3}$ K at a
frequency of 2 GHz would place useful limits on current models of the
ionized gas distribution.

In these calculations, we have assumed $T_e = 2 \times 10^4$ K,
consistent with measurements of the Ly$\alpha$ forest at $z \sim
2.4$--$3.9$ (\cite{Zaletal01} 2001). The free-free emission does not
depend strongly on $T_e$ ($\epsilon_\nu \propto T_e^{-0.35}$,
including the Gaunt factor), so allowing the temperature to vary as a
function of density, halo mass, or redshift only leads to a reduction
in the mean brightness by a few tens of percent or less.  Thus, we do
not expect uncertainties in the thermal history of the ionized gas to
be a significant limitation in these estimates.
%SF, removed following for space reasons
% (because the
%fractional contribution from large halos is small).  Additionally, the
%mean IGM temperature is likely to be higher near reionization at $z
%\sim 10$ (\cite{HuiHai03} 2003). Such corrections are again minor
%because $T_\nu$ is dominated by clumped ionized gas at $z \la 3$ and
%not by electrons at $z \sim 10$.  

\section{Angular Fluctuations from Free-free Emission}

We now consider the observability of spatial fluctuations in the
free-free intensity. In general, fluctuations in the brightness
temperature of the free-free background arise from fluctuations in the
electron density field.  For simplicity, we will neglect anisotropies
generated by fluctuations in the ionized fraction during reionization,
as the fractional contribution to the mean intensity is dominated by
ionized gas at $z \la 3$.

\begin{figure}[t]
\psfig{file=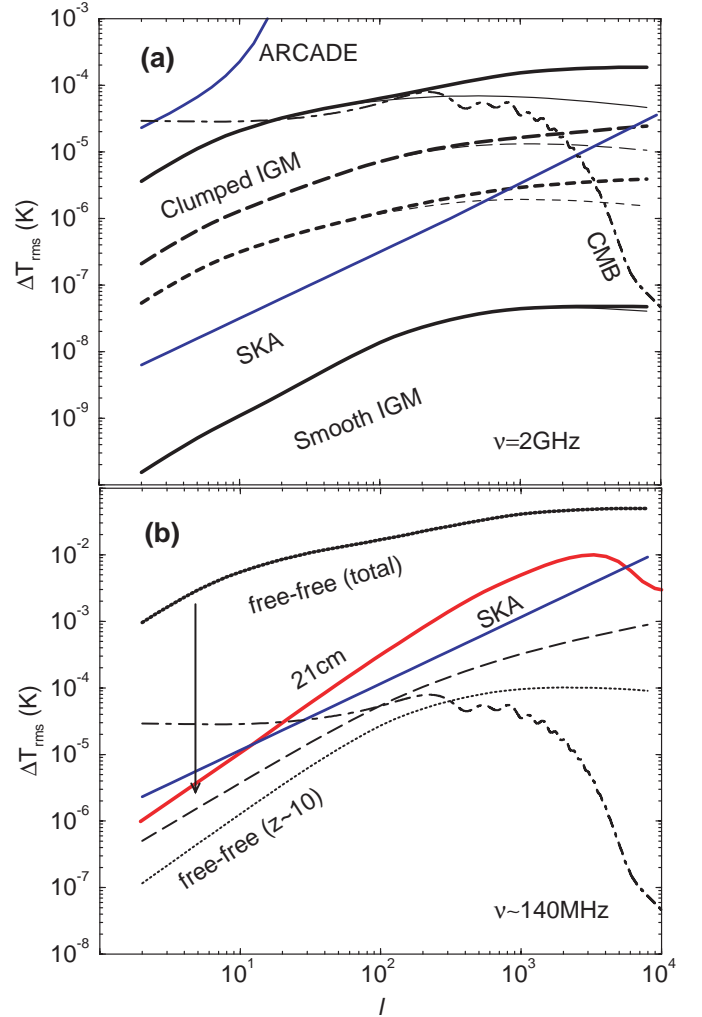,width=3.5in,angle=0}
\caption{(a) The rms brightness temperature fluctuations of the
free-free background at $\nu_{\rm obs}=2$ GHz, where $\Delta T_{\rm
rms} = \sqrt{l(l+1)/2 \pi C_l}$. The curves refer to the clumping
models in Fig.~1.  The thin lines are for fluctuation power based on
equation~(2) using the bias factor in equation~(3) and the linear
power spectrum, while the thick lines include non-linearities in the
dark matter density field following \cite{Smietal03} (2003). At $l \ga
10^3$, non-linearities are important and our estimate for fluctuation
power spectrum is likely to be only approximate given uncertainties in
the spatial distribution of ionized clumps within each halo. At these
small scales, we may also have underestimated power as we do not
consider the Poisson contribution related to the finite number density
of free-free sources.  For reference, we also show primordial CMB
anisotropies as well as noise power spectra for ARCADE and for the
SKA, assuming one month of continuous observations for the latter.
(b) The brightness temperature fluctuations at $\nu_{\rm
obs}=140$~MHz. Here, we show the 21 cm signal in a narrow window
around $z=10$ (taken from \cite{Zaletal03} 2003), the CMB, and
free-free fluctuations.  The dotted line shows the free-free
fluctuations including only those halos in the redshift interval of
the 21 cm power spectrum; the long-dashed line shows the
(absolute value of the) cross-correlation between the 21 cm and
free-free signals in this redshift window.  }
\label{fig:cl}
\end{figure}

We define the angular power spectrum of the free-free background to be
$\langle T_b(\vecl) T_b(\vecl') \rangle = (2\pi)^2
\delta(\vecl+\vecl') C_l$, , where $T_b(\vecl)$ is the Fourier
transform of free-free brightness temperature.  Using the Limber
approximation (\cite{Lim54} 1954), we can write the brightness
temperature anisotropy power spectrum, observed at a frequency $\nu$,
as
\begin{eqnarray}
&&C_l^\nu = \frac{c^4}{4 \nu_{\rm obs}^4 k_B^2} \int
  \frac{d\chi}{\da^2}  \left[\frac{\epsilon_\nu(\chi) x_e(z)^2
      C(z)}{(1+z)^4}\right]^2  
P_{e e}\left(k=\frac{l}{\da},\chi\right) \, , \nonumber \\
\end{eqnarray}
where $\da$ is the comoving angular diameter distance and $P_{ee}(k)$
is the three-dimensional power spectrum of the electron field.  To
calculate this quantity we make use of the halo approach to large
scale structure clustering (\cite{CooShe02} 2002). We assume that on
the scales of interest (from degrees to arcminutes), we can describe
fluctuations in the density field through the 2-halo part of the power
spectrum, which includes the clustering of different ionized
halos. Thus we write $P_{ee}(k) = b_{ee}^2 P^{\rm lin}(k)$ in terms of the
linear power spectrum. We assume that the ionized patches exist in all
halos with a minimum mass, $M_{\rm min}(z)$ that corresponds to
$T_v=10^4$~K and calculate the bias factor of these halos with respect
to the linear density field.  Thus
\begin{equation}
b_{ee}(z) = \frac{\int_{M_{\rm min}}^\infty M^2 b_{\rm halo}(M,z)
  dn/dM}{\int_{M_{\rm min}}^\infty M^2 dn/dM} \, , 
\end{equation}
where $b_{\rm halo}(M,z)$ is the bias factor of the dark matter halos
(\cite{Moetal97} 1997).  Note the additional $M^2$ weighting, which
takes into account the fact that the relevant bias factor is that of
$n_e^2$.  Essentially, we assume that the free electrons are in the
form of ionized clumps in each dark matter halo, and we assign a halo
occupation number proportional to the dark matter halo mass (in
analogy to the halo approach for galaxy biasing).  With this
weighting, $b_{ee}$ exceeds the dark matter bias, although the
difference is minor at redshifts of a few.

Our results for the angular fluctuations are summarized in Fig.~2(a).
The typical fluctuations are $\la 10^{-4}$~K at 2 GHz, a few percent
of the mean brightness. This is consistent with a mildly biased field
that traces the large scale dark matter field and has its dominant
contribution at redshifts of a few.  Again the \cite{Oh99} (1999)
model predicts substantially larger fluctuations than the other
models, suggesting that measurements may distinguish between the two
predictions (though without constraining the details of the large
scale IGM clumping).  For comparison, we also show the noise level, at
each multipole, expected for the ARCADE experiment, if it attempts to
make anisotropy measurements with a noise per beam size pixel of 100
$\mu$K and a beam size of 16$\arcdeg$, and for the planned Square
Kilometer Array (SKA). In calculating the SKA noise, we follow the
approach of \cite{Zaletal03} (2003) and assume a system temperature
$T_{\rm sys}=50$ K at 2 GHz, a bandwidth of 10 MHz, and continuous
observations over 4 weeks. We also show the anisotropy power spectrum
of primordial CMB fluctuations.  While the CMB fluctuations swamp
those expected from the free-free background, the CMB contamination
can be reduced through two methods: (1) because free-free fluctuations
are dominated by point-like clumped regions in the IGM, which are
likely to be associated with dark matter halos, one can filter the
large scale fluctuations of CMB and concentrate on the remaining
point-like emission, or (2) one can use existing cosmic-variance
limited CMB maps at higher frequencies to clean the CMB
contamination. We see that SKA can in principle detect free-free
fluctuations on scales $l \la 3000$.  The main source of confusion
will be the uncertain contribution from extragalactic radio sources,
such as radio galaxies (\cite{DiMetal02} 2002). Because these sources
are also point-like, separating them from free-free fluctuations will
require frequency information.

To estimate the background for 21 cm measurements, we consider
free-free fluctuations at $\nu_{\rm obs}=140$ MHz, corresponding to
line emission from neutral hydrogen at $z \approx 10$. We summarize
our results in Fig.~2(b).  We also show the SKA noise power spectrum,
where we have set $T_{\rm sys}=200$ K and the bandwidth to 0.2 MHz,
because one wants narrow redshift slices in order to observe many
independent regions for 21 cm fluctuations.  The free-free
fluctuations are comparable to or greater than the rms fluctuations in
the 21 cm background, as estimated by \cite{Zaletal03} (2003), while
contamination by the CMB is negligible.
%SF
As noted by \cite{OhMac03} (2003), free-free fluctuations present a
substantial foreground; however, the free-free spectrum is
smooth in frequency space. while the 21 cm signal varies strongly over
$\sim 0.2$ MHz.  Thus multi-frequency differencing can efficiently remove
the smooth background (\cite{Zaletal03} 2003), but only so long as the
two signals are uncorrelated.
%SF, end
We expect this to be the case with free-free emission: halos
with ionized gas will emit free-free radiation but not 21 cm
radiation, so the two fields are strongly anti-correlated.

To test how important this effect is, we show the free-free emission
from halos in the same redshift slice as the 21 cm signal (a window of
0.2 MHz about $z=10$) with the dotted line.  To construct this curve
we have used the \cite{Oh99} (1999) clumping model; this gives a mean
ionizing rate roughly similar to the simulations on which the simple
estimate of \cite{Zaletal03} (2003) are based (\cite{Sok03} 2003).
Note that $\Delta T_{\rm rms} \propto x_e^2 C(z)$ and can thus be
easily scaled for different clumping scenarios.  In particular, in the
\cite{Oh99} (1999) approach the free-free signal will just be
proportional to $\dot{n}_{\rm ion}$.  We then estimate the cross-power
spectrum between the free-free and 21 cm signals by assuming that they
are perfectly anti-correlated on large scales; the (absolute value of
the) result is shown by the long-dashed line in the Figure.  This
curve is thus our estimate for the \emph{maximally} cleaned 21 cm map;
it shows that the two are correlated only on the percent level.
Because it is smaller than the noise power spectrum for SKA
observations, the free-free signal can be efficiently removed with
multi-frequency data.

\section{Conclusions}

To summarize, we find that the cumulative specific intensity of the
free-free background at low radio frequencies is strongly sensitive to
the spatial clumping of free electrons in the IGM. Using a variety of
existing models for the clumping of ionized gas and its redshift
evolution, we find that the expected free-free distortion to the CMB
blackbody spectrum, at frequencies of order a few GHz and below, is
within the reach of upcoming experiments such as ARCADE.  The dominant
contribution to the background is from ionized gas at $z \sim 3$,
because the clumping factor is generally large below this redshift.
The free-free background varies across the sky at the level of a few
to at most ten percent.  These fluctuations can be detected with
experiments such as the SKA, though a careful separation of CMB
anisotropies and other low-frequency radio point sources will be
required.  At frequencies of order 200 MHz, where observations of 21
cm radiation from neutral hydrogen during and before the epoch of
reionization are planned, angular fluctuations from free-free emission
could become a significant source of confusion.  However, we have
shown that the cross-correlation between the 21 cm and free-free
signals is small, suggesting that the free-free foreground can be
removed to good accuracy.
%This contamination,
%however, can be reduced significantly through frequency differencing
%(\cite{Zaletal03} 2003; \cite{MorHew03} 2003), but only so long as the
%free-free and 21 cm signals are uncorrelated.  Fortunately, we
%estimate this cross-correlation to be small: at $z\sim10$, it is less
%than a few percent for moderate neutral fractions.  This is
%substantially below the expected level of 21 cm fluctuations,
%suggesting that low frequency measurements will not be limited by
%free-free emission from the IGM.

\acknowledgements

This work is supported by the Sherman Fairchild foundation and DOE
DE-FG 03-92-ER40701 (AC). We thank A. Benson and Z. Haiman for
information related to their clumping models, S.~P. Oh for useful
comments on the manuscript, and A. Kogut for ARCADE specifications.

\end{document}